\documentclass[preprint,showpacs,preprintnumbers,amsmath,amssymb,
aps,prl]{revtex4}
\usepackage{amsmath,amsfonts,amssymb,amscd,amsthm,epsf,graphicx}
\usepackage[english]{babel}
\usepackage{times}
\usepackage[latin1]{inputenc}
\usepackage{comment}

\newcommand{\mR}{\mathbb{R}}
\newcommand{\mT}{\mathbb{T}}

\newcommand{\mZ}{\mathbb{Z}}

\usepackage{graphicx}

\newcommand{{\fhi}}{\varphi}

\newcommand{\rmcite}[1]{{\rm{\cite{#1}}}}

\begin{document}

\title{Strange Nonchaotic Attractors in Harper Maps}
\author{\`Alex Haro$^1$ and Joaquim Puig$^2$}
\affiliation{
$^1$Dept. de Matem\`atica Aplicada i An\`alisi, 
Universitat de Barcelona, Gran Via 585, Barcelona 08007, SPAIN 
\email{haro@mat.ub.es} 
\\
$^2$Dept. de Matem\`atica Aplicada I,
Universitat Polit\`ecnica de Catalunya, Diagonal 647, Barcelona 08028, SPAIN
\email{joaquim.puig@upc.edu}
}
\thanks{A.H. supported by INTAS project 00-221 and 
MCyT/FEDER Grant BFM2003-07521-C02-01. J.P supported by 
MCyT/FEDER Grant BFM2003-09504-C02-01 and CIRIT 2001 SGR-70.}


\begin{abstract}
{We study the existence of Strange Nonchaotic Attractors (SNA) in 
the family of Harper maps, proving that they are typical but not 
robust in this family.
Our approach is based on the theory of linear 
skew-products and the spectral theory of   
Schr\"odinger operators.
}
\end{abstract}

\maketitle

The study of the attractors of a dissipative dynamical system is a topic
of great interest, because 
these invariant sets trap the evolution of a 
large subset of the phase space and capture the asymptotic behavior. 
It has been known for a long time that 
attractors can be {\em strange} \cite{RuelleT71}, i.e.
geometrically complicated. 
The first examples of strange attractors were {\em chaotic}, i.e.  
with dependence sensitive on initial conditions
\cite{EckmannR85}. 
In \cite{GrebogiOPY84} were found strange attractors that are nonchaotic,
and it has stimulated much numerical experimentation
(see the review \cite{prasad-etc:sna}) as well as rigorous analysis 
\cite{Keller96,Glend02,Stark03}.

Our interest in this Letter is to show the existence and abundance of 
{\emph{Strange Nonchaotic Attractors}} (SNA for short) in the family of 
Harper maps. This is a family of 1D quasi-periodically forced maps
that many authors have suggested as a scenario in which SNA appear 
\rmcite{ketoja-satija,ketoja-satija:self,negi-ramaswamy:collision,
negi-ramaswamy:critical,mestel-osbaldestin-winn}\rmcite{bondeson-etc}.
By SNA we mean here 
an invariant set that is a graph of a measurable and  
nowhere continuous function (it is \emph{Strange}), that carries a 
quasi-periodic dynamics (it is \emph{Nonchaotic}) and it attracts 
exponentially fast almost every orbit in phase space (it is an 
\emph{Attractor})
\footnote{The definition itself of SNA is the subject of much debate and the 
definition given here may not be suitable for other models.}.
We prove that these SNA are typical but 
not robust in the family of Harper maps, 
in the sense that they exist for a positive measure Cantor set 
of the parameter space.  

In our analysis, we exploit the connections between 
(a) the dynamical properties of the 
\emph{Harper map} (a 1D quasi-periodically forced map); 
(b) the spectral properties of the \emph{Harper operator}
(an example of a quasi-periodic Schr\"odinger operator);
(c) the geometrical properties of the \emph{Harper linear skew-product} 
(a 2D quasi-periodically forced linear map).

In recent years 
our knowledge of the spectral properties of the Harper operator,
also  known as the Almost Mathieu operator, 
and related quasi-periodic Schr\"odinger operators
has advanced spectacularly. The progress made will be 
relevant to our approach. In particular, the connections between 
(b) and (c) have been successfully applied to 
the solution of the ``Ten Martini Problem'' \cite{puig}, 
on the Cantor structure of the spectrum of the Harper operator.

The connection between (a) and (c) in similar models 
has been used to study the linearized 
dynamics around invariant tori in quasi-periodic systems \cite{HLlex}.
Specifically, the formation of SNA in this linearized dynamics 
is suggested to be a mechanism of breakdown of invariant tori
\cite{HLlverge}.

A consequence of our approach is that 
neither arithmetic properties of the frequency 
of the quasi-periodic forcing
nor localization properties of the spectrum of the Harper operator are 
crucial for the existence of SNA.
  
The family of quasi-periodically forced dynamical systems under investigation 
in this Letter is the family of \emph{Harper maps}  
\begin{equation}\label{eq:harpermap}
\begin{array}{ccl}
y_{n+1} &=& 
\underbrace{\frac{1}{a-b\cos{(2\pi\theta_n)}-{y_n}}}_{f_{a,b}({y_n},\theta_n)}
,
\\
\theta_{n+1} &=& \theta_n + \omega  \qquad (\mathrm{mod~} 1), 
\end{array}
\end{equation}
where $y \in \overline{\mR}= [-\infty,+\infty]$ 
and $\theta\in \mT= {\mR}/{\mZ}$  are the phase space 
variables, $a,b$ are 
the parameters, and $\omega$ is the \emph{frequency} (it is assumed to be 
irrational).   

Notice that a Harper map is a skew-product map  
\[
F_{a,b,\omega}(y_n,\theta_n) = (f_{a,b}(y_n,\theta_n),\theta_n + \omega),
\]
defining a dynamical system in $\overline{\mR}\times\mT$ 
whose evolution from an initial condition 
$(y_0,\theta_0)$
is described by the $Nth$-power
$F^{(N)}_{{a,b,\omega}}(y_0,\theta_0) = 
(f^{(N)}_{a,b}(y_0,\theta_0),\theta_0 + N\omega)$, for $N\in \mZ$.

In a Harper map the parameter $a$ is called the \emph{energy} or the 
\emph{spectral parameter} because after writing ${y}_n=x_{n-1}/x_n$ this 
family is equivalent to the family of \emph{Harper equations}, which are 
second-order difference equations
\begin{equation}\label{eq:harper}
 x_{n+1} + x_{n-1} + b\cos\left(2\pi(\theta_0 + n \omega)\right)x_n= a x_n.
\end{equation}
These equations are physically relevant because they show up as 
eigenvalue equations of 
 \emph{Harper operators} (also known as \emph{Almost Mathieu operators}),
\begin{equation}\label{eq:almostmathieu}
\left(H_{b,\omega,\theta_0} x\right)_n= 
x_{n+1} + x_{n-1} +  b\cos\left(2\pi(\theta_0+n\omega)\right)x_n.
\end{equation}
These are  bounded and self-adjoint operators on $l^2(\mZ)$ whose spectrum,
that does not depend on $\theta_0$, describes the 
energy spectrum of an electron in a rectangular lattice subject to a 
perpendicular magnetic flux \rmcite{harper,sokoloff}. 

The formulation of the second-order difference equation
\eqref{eq:harper} as a first-order system is the 
\emph{Harper linear skew-product}
\begin{equation}\label{eq:harperskew}
\begin{array}{ccl}
\underbrace{\left(\begin{array}{c}
x_{n+1} \\
x_{n} 
\end{array}\right)}_{v_{n+1}} &=&
\underbrace{\left(
\begin{array}{cc}
a - b \cos\left(2\pi\theta_n\right)  & \;  -1 \\
1 & \; \phantom{-} 0
\end{array}
\right)}_{M_{a,b}(\theta_n)}
\underbrace{\left(\begin{array}{c}
x_{n} \\
x_{n-1} 
\end{array}\right)}_{v_n}, \\ 
\theta_{n+1} &=& \theta_n + \omega \qquad (\mathrm{mod~} 1), 
\end{array}
\end{equation}
 whose evolution is given by the \emph{Harper cocycle} 
\begin{equation}\label{eq:transfermatrix}
M^{(N)}_{a,b,\omega}(\theta_0)= 
\begin{cases}
M_{a,b} \left(\theta_{N-1}\right)\dots M_{a,b}(\theta_0)  
& \text{if $N>0$,} \\ 
I  & \text{if $N= 0$,} \\ 
M_{a,b}^{-1} \left(\theta_N\right)\dots M_{a,b}^{-1}(\theta_{-1})  
& \text{if $N<0$.}
\end{cases} 
\end{equation}

Note that (\ref{eq:harpermap}) describes the evolution of the slope 
$y_n$ of vectors $v_n$ under the action of the linear skew-product 
(\ref{eq:harperskew}). That is, (\ref{eq:harpermap}) is the 
\emph{projectivization} of (\ref{eq:harperskew}). 

To understand the dynamics of Harper linear skew-products 
it is important to know the growth properties of 
the solutions. The exponential growth is measured by the  
\emph{Lyapunov exponents} which we now define. Given any nontrivial 
initial condition of the skew-product (\ref{eq:harperskew}), 
$(v_0,\theta_0)$ with $v_0\neq 0$, the (forward) Lyapunov exponent for 
$(v_0,\theta_0)$ is the limit
\begin{equation}\label{eq:lyapunovindividual}
\begin{split}
\lambda_{a,b,\omega}(v_0,\theta_0) 
& = 
\lim _{N \to +\infty} 
\frac{1}{N}\log{\left|v_N\right|},
\\
& =
\lim _{N \to +\infty} 
\frac{1}{N}\log{\left| M_{a,b,\omega}^{(N)}(\theta_0) v_0\right|}
\end{split}
\end{equation}
whenever the limit exists (in which case it is finite). 
If $v_0=(x_{-1},x_0)$ and ${y}_0=x_{-1}/x_0$ then one can also define the 
(forward) Lyapunov exponent of the Harper map (\ref{eq:harpermap}) 
for the initial condition $({y}_0,\theta_0)$ by 
\begin{equation}\label{eq:lyapunovharpermap}
\begin{split}
{\beta}_{a,b,\omega}({y}_0,\theta_0) &= 
\lim _{N \to +\infty} 
\frac{1}{N}\log{\left|\frac{\partial {y}_N}{\partial {y}_0}({y}_0,\theta_0)\right|} \\
&= \lim _{N \to +\infty} \frac{1}{N}\log{
 \left| m^{(N)}_{a,b,\omega}(y_0,\theta_0)
 \right|
 }
\end{split}
\end{equation}
where
\[
   m^{(N)}_{a,b,\omega}(y_0,\theta_0) = 
  \frac{\partial f_{a,b}}{\partial {y}} \left({y}_{N-1},\theta_{N-1}\right)\dots 
  \frac{\partial f_{a,b}}{\partial {y}} \left({y}_{0},\theta_{0}\right) \ .
\]
An easy computation shows the relation
\[
\beta_{a,b,\omega}({y}_0,\theta_0)= -2 {\lambda}_{a,b,\omega}(v_0,\theta_0).
\]
Backward Lyapunov exponents are defined by replacing 
$\lim_{N\to+\infty}$ with $\lim_{N\to-\infty}$ in the above formulation.

Oseledec \rmcite{Oseledec68}
showed that for almost every initial condition $(v_0,\theta_0)$
the Lyapunov exponent exists and equals 
the \emph{averaged Lyapunov exponent}
\[
 \overline{\lambda}_{a,b,\omega}= 
 \lim _{N \to +\infty} 
 \frac{1}{N}\int_{\mT}\log{\left| M^{(N)}_{a,b,\omega}(\theta)\right|} d\theta,
\]
which is never negative and exists by the Kingman subadditive ergodic 
theorem \rmcite{kingman}.

The case of the nonzero averaged Lyapunov exponent,  
$\overline{\lambda}_{a,b,\omega}>0$, which we call \emph{hyperbolic}, is 
important for our purposes. 
In this case, there exists a full measure set 
$\Theta\subset\mT$ such that for every $\theta \in \Theta$  
one has a splitting 
\begin{equation}\label{eq:splitting}
\mR^2= W^s(\theta) \oplus W^u(\theta)
\end{equation}
characterized by 
\begin{equation}\label{eq:definitionws}
v \in W^s(\theta)\setminus\{0\} 
\quad \Leftrightarrow \quad 
\lim_{N\to \pm\infty} \frac{1}{N} \log{\left| M^{(N)}(\theta) v \right|}= 
-\overline{\lambda} 
\end{equation}
and 
\begin{equation}\label{eq:definitionwu}
v \in W^u(\theta)\setminus\{0\} 
\quad \Leftrightarrow \quad 
\lim_{N\to \pm\infty} \frac{1}{N}\log{\left| M^{(N)}(\theta) v \right|}= 
+\overline{\lambda}. 
\end{equation}
$W^s(\theta)$ and $W^u(\theta)$ are the 
\emph{stable} and \emph{unstable} subspaces at $\theta$, respectively.
The elements of the set $\Theta$ are referred to as the \emph{Lyapunov 
regular points}.

In the phase space $\mR^2\times\mT$, one can form the product sets 
$W^s$ and $W^u$ whose elements are pairs $(v,\theta)$ with 
$v \in W^{s}(\theta)$ or  $W^{u}(\theta)$ respectively 
(whenever these subspaces are defined). 
These are the \emph{stable} and \emph{unstable subbundles}. 
According to Oseledec \rmcite{Oseledec68} the $\theta$-dependence 
of the decomposition is measurable but not necessarily continuous. 

When the splitting \eqref{eq:splitting} is defined 
for \emph{all} $\theta$, 
that is $\Theta= \mT$ (hence $\theta$-dependence of the subbundles 
is continuous
\footnote{In fact Johnson \& Sell \rmcite{johnson-sell} and Johnson 
\rmcite{johnson:analyticity} prove that in such a case 
the decomposition is as regular as the 
original system, in the case of Harper map, real analytic, 
whenever it is continuous. 
See also  \rmcite{HirschPS77,trilogy1}.\label{foo:continuity}}) 
the linear skew-product  
is said to be \emph{uniformly hyperbolic}.
Otherwise it is said to be \emph{nonuniformly hyperbolic}. 

In the Harper map, we can determine whether or not hyperbolicity 
is uniform by
looking at the spectral problem of (\ref{eq:almostmathieu}). 
Indeed,  an energy $a$ is in the spectrum of 
the Harper operator (\ref{eq:almostmathieu}) if, and only if, the corresponding
linear skew-product (\ref{eq:harperskew}) is \emph{not} uniformly hyperbolic 
\rmcite{mane78,johnson:recurrent}. We will use an implication of this result: 
if $a$ is in the spectrum of the Harper operator and the averaged 
Lyapunov exponent is nonzero at $a,$ then the linear skew-product 
is nonuniformly hyperbolic. Let us now see that in this  case the 
corresponding Harper map has a SNA.

The above concepts of hyperbolicity can be translated to the dynamics of 
the Harper map \eqref{eq:harpermap} 
(which reflects how the linear skew-product \eqref{eq:harperskew} 
changes directions of vectors). 
Recall that, in the hyperbolic case,  ${\beta}<0$, 
there exist two invariant subbundles $W^s$ and $W^u$, 
for $\theta \in \Theta$,  which are measurable as a function of 
$\theta$ and satisfy (\ref{eq:definitionws}) and (\ref{eq:definitionwu}). 
We define  ${y}^s(\theta)$ and ${y}^u(\theta)$  as the slopes of the 
subbundles: they are the only elements of ${\overline{\mR}}$ such that 
$(1,{y}^s(\theta))^T$ and $(1,{y}^u(\theta))^T$ belong to $W^s(\theta)$ 
and $W^u(\theta)$ respectively. The product sets
\[
Y^s= \left\{ ({y}^s(\theta), \theta), \;\theta \in \Theta\right\}
\text{ and }
Y^u= \left\{ ({y}^u(\theta), \theta), \;\theta \in \Theta\right\}
\]
are invariant under the Harper map and have quasi-periodic dynamics;
thus ${Y}^s$ and ${Y}^u$ are \emph{nonchaotic} invariant sets. 

Still in the hyperbolic case, the decomposition of $\mR^2$ 
into direct sum of $W^s(\theta)$ and $W^u(\theta)$, $\theta \in \Theta$,  
implies that every pair $(v,\theta)$ of this skew-product 
(\ref{eq:harperskew}) other than the stable subbundle, is attracted to 
the unstable subbundle and grows exponentially  in norm. 
Looking at directions (which is what the Harper map retains), 
forward orbits with initial condition $({y},\theta)$ 
(other than $({y}^s(\theta),\theta)$) are exponentially attracted to 
${Y}^u$, that is
\begin{equation}\label{eq:convergencetosna}
\lim_{N\to+\infty}\frac{1}{N}\log\left|f_{a,b,\omega}^{(N)}({y},\theta)- 
{y}^u(\theta_N)\right| = 
\beta = -2\overline{\lambda} < 0,
\end{equation}
while backward orbits (other than $({y}^u(\theta),\theta)$) 
are exponentially attracted to 
${Y}^s$. Thus ${Y}^u$ is a \emph{nonchaotic attractor} for the Harper map: 
for almost every initial condition, orbits are exponentially attracted to it.
Similarly $Y^s$ is a \emph{nonchaotic repellor}.

Let us now relate the uniformity of hyperbolicity in the skew-product to the 
regularity of these nonchaotic attractors. If the skew-product 
\eqref{eq:harperskew} is uniformly hyperbolic then 
the invariant subbundles $W^s$, $W^u$ are defined in all $\mT$ and are 
continuous, and so are their projectivizations ${y}^u$ and ${y}^s$.  

In contrast, if the skew-product is nonuniformly hyperbolic, then the 
invariant subbundles are measurable but not continuous and their 
projectivizations ${y}^u$ 
and ${y}^s$ are measurable but not continuous functions of $\theta$.  
Moreover, discontinuities are propagated by the quasi-periodic dynamics and 
the invariance property of the attractor: if ${y}^u$ is discontinuous at a 
single $\theta_0$ then the same happens for $\theta_N= \theta_0+N\omega$ 
for all $N \in \mZ$, so that the function ${y}^u$ is \emph{nowhere continuous}.
The same result happens for ${y}^s$. 

In summary, in the nonuniformly hyperbolic case, we will say that $Y^u$ is 
a \emph{Strange Nonchaotic Attractor (SNA)} of the Harper map 
because the following properties are satisfied:
\begin{enumerate}
\item[(i)] $Y^u$ is the graph of a measurable function of $\theta$,
$y^u$, which is nowhere continuous ($Y^u$ is \emph{Strange});
\item[(ii)] $Y^u$ is an invariant set of the Harper map with 
quasi-periodic dynamics ($Y^u$ is \emph{Nonchaotic});
\item[(iii)] Almost every orbit in phase space is attracted to 
${Y}^u$ at exponential rate ($Y^u$ is an \emph{Attractor}).
\footnote{In principle one could look for SNA without exponential rate 
of attraction (for instance given by a power law). A candidate for 
such objects would be a Harper map at critical coupling $b=2$ and energies 
in the spectrum, as discussed in \rmcite{keller-etal}. 
For definiteness, we focus on the exponential case.}
\end{enumerate}

The existence of nonuniformly hyperbolic linear skew-products 
was already shown by Herman \rmcite{herman}, who proved that 
\begin{equation}\label{eq:hermanbound}
\overline{\lambda}_{a,b,\omega} \ge \max\left(0, \log\frac{|b|}{2} \right)
\end{equation}
as long as $\omega$ is irrational.  Moreover, Bourgain \& Jitomirskaya 
\rmcite{bourgain-jitomirskaya} prove that the equality in 
\eqref{eq:hermanbound} holds if, and only if, $a$ is in the spectrum of the 
Almost Mathieu operator. 

Thus, for $|b|>2$ and $\omega$ irrational, a Harper map has a SNA if, and 
only if, $a$ belongs to the spectrum. Since the 
measure of the spectrum is given by the formula $|4-2|b||$ 
\rmcite{jitomirskaya-krasovsky} these SNA are \emph{persistent in measure} 
in the family of Harper maps. They are not, however, persistent in open sets, 
since the spectrum is a Cantor set
\rmcite{choi-elliott-yui,puig,avila-jitomirskaya}, and therefore any SNA in 
a Harper map can become a regular attractor by means of an 
arbitrarily small perturbation.
As an example, using the symmetry of the spectrum, the Harper map has 
a SNA for any $|b|>2$, $\omega$ irrational and $a=0$ 
(this value always belongs to the 
spectrum).

\begin{figure} 
\includegraphics[width=\linewidth]{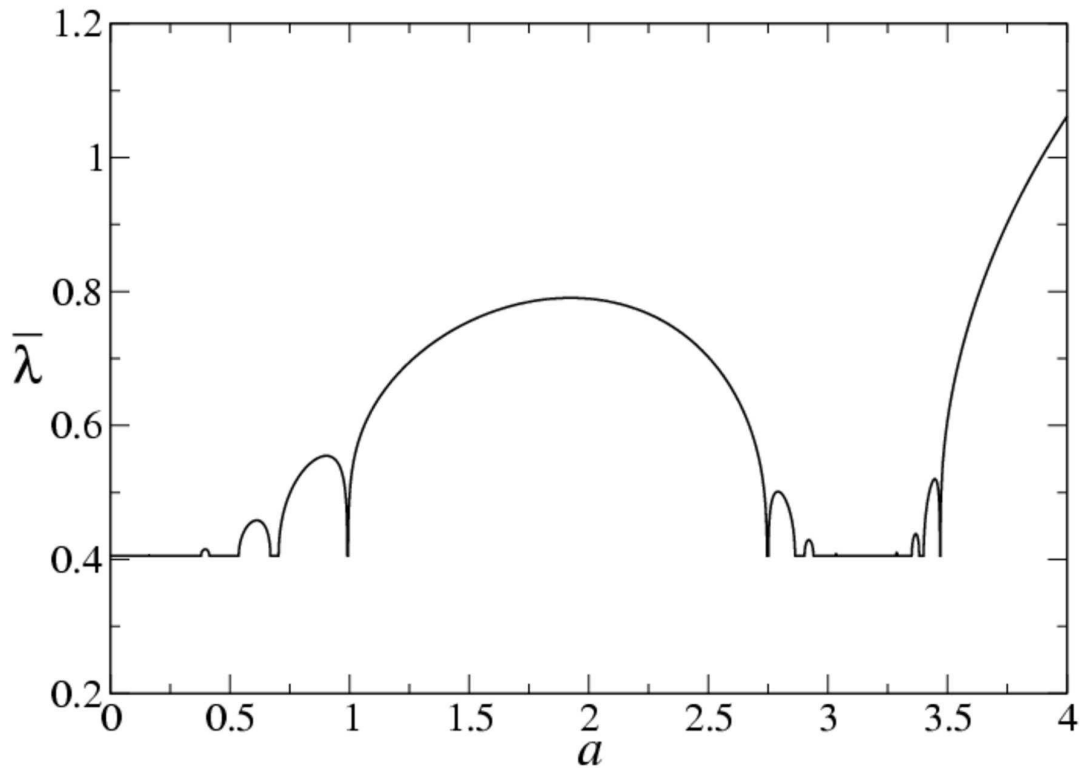}
\caption{\label{fig:lyapb-3} 
The averaged Lyapunov exponent as a function of $a$.}
\end{figure}

As an illustration of the above rigorous results, we performed 
several numerical computations. In the following, we chose the 
irrational frequency $\omega= \displaystyle \frac{e}{4}$ 
\footnote{Most of the authors in the literature choose 
the golden mean $\frac{1}{2}(\sqrt{5} -1)$
as the frequency. This number has very 
specific arithmetical properties. In particular it has a periodic 
(in fact, constant) continuous fraction expansion which makes it
convenient for renormalization procedures. 
The relative distance of $\omega= \displaystyle \frac{e}{4}$ 
to the golden mean is less than 10\%, but their arithmetical properties are 
very different. Our choice aims to emphasize that the results presented 
here are independent of the arithmetical properties
of the frequency \cite{HLlex,HLlverge}. 
}
,
$b= 3$, and we considered $a$ 
as a moving parameter. The averaged Lyapunov exponent as a function 
of $a$ is displayed in Figure~\ref{fig:lyapb-3}. Notice that 
\eqref{eq:hermanbound} implies that $\overline\lambda_a \geq \frac{\log 3}{2}$, 
so that the Harper cocycle is hyperbolic for all the values $a$. Moreover, 
the equality holds only if the cocycle is nonuniformly hyperbolic. 
As a result, the values of $a$ for which $Y^u$ is a SNA of the Harper map 
correspond to the ``flat pieces'' of the graph in Figure~\ref{fig:lyapb-3},
which lie in  a Cantor set  of measure 2. The ``bumps'' appear in gaps of the 
spectrum, that is energies $a$ in the resolvent set, for which 
$Y^u$ is an invariant attracting continuous curve. 
Hence, gaps are labelled by a topological index 
which is the number of turns of $Y^u$ on $\overline\mR$ \cite{johnson-moser}.

\begin{figure}
\begin{tabular}{cc}
\hspace{-0.05\linewidth}
\includegraphics[width=0.55\linewidth]{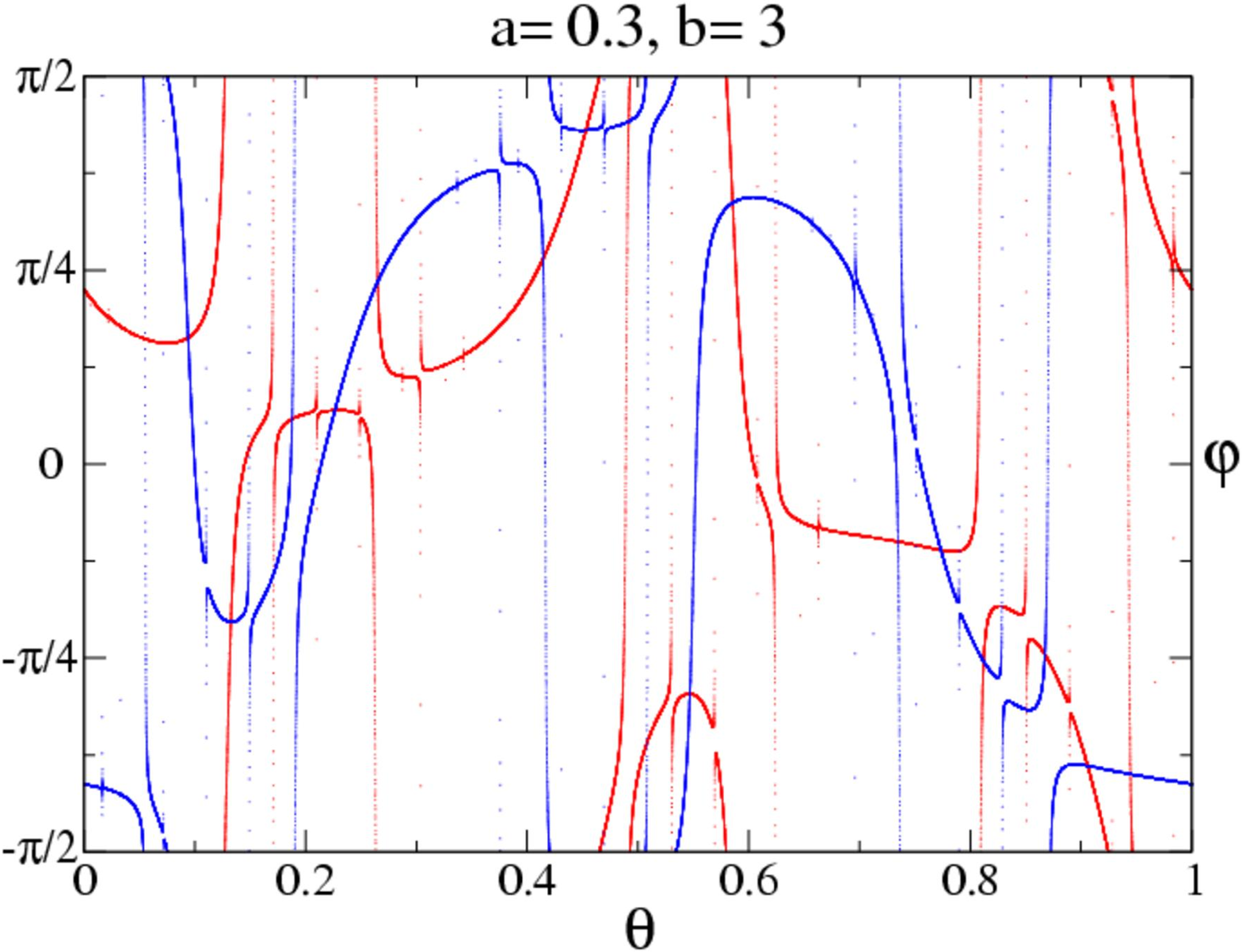} &
\hspace{-0.05\linewidth}
\includegraphics[width=0.55\linewidth]{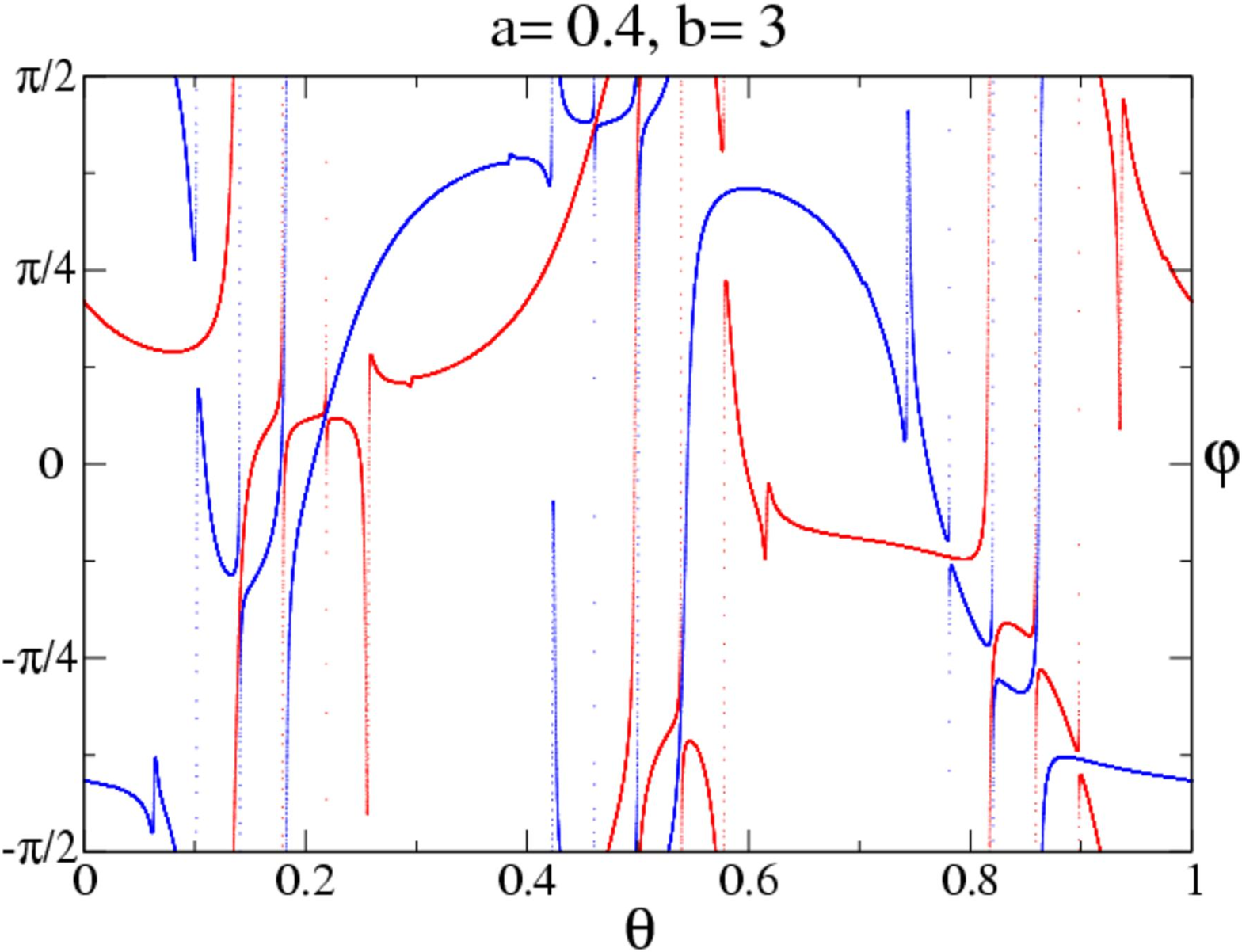} 
\\
\hspace{-0.05\linewidth}
\includegraphics[width=0.55\linewidth]{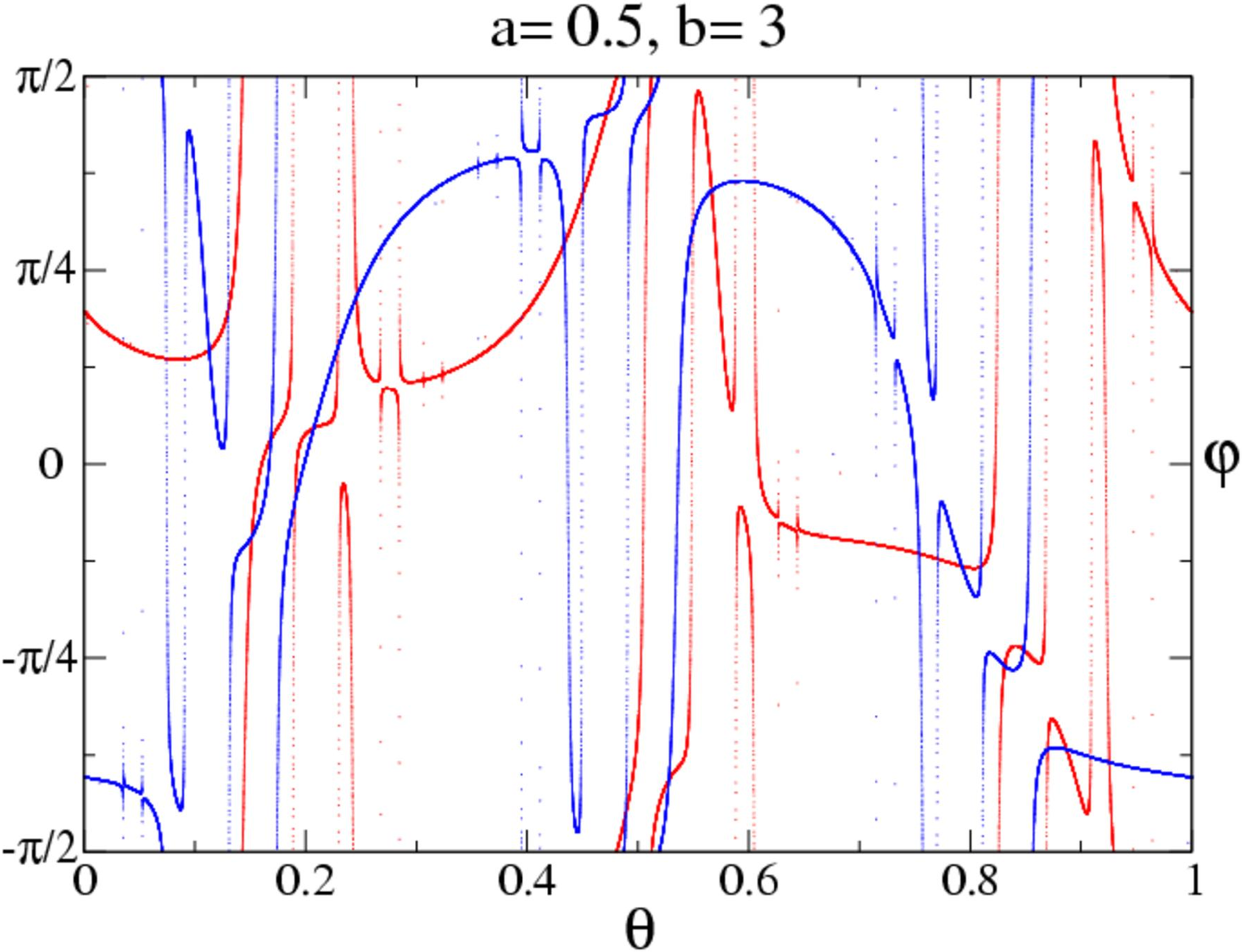} &
\hspace{-0.05\linewidth}
\includegraphics[width=0.55\linewidth]{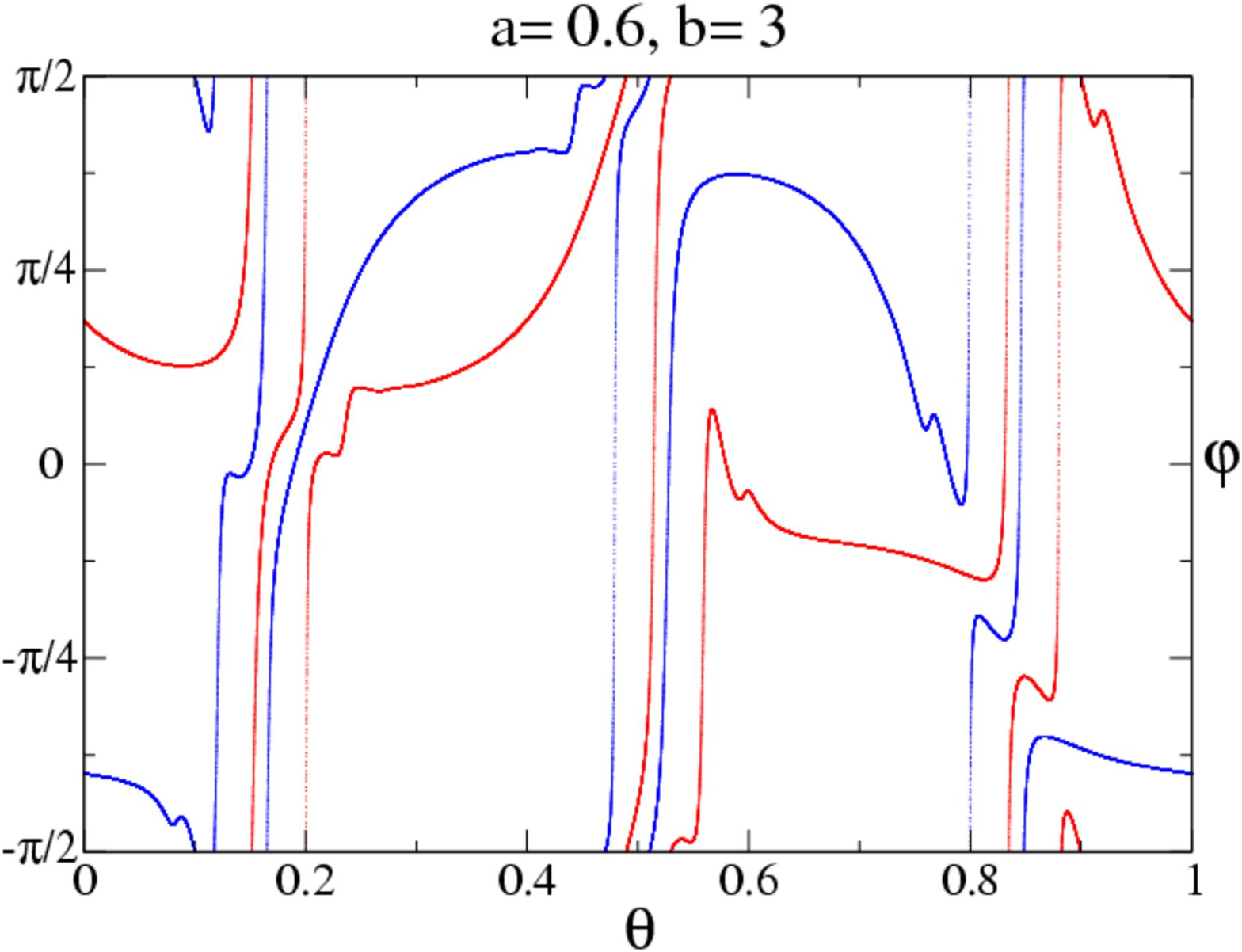}
\end{tabular}	
\caption{\label{fig:harperb-3a}
The attractor $Y^u$ and the repellor $Y^s$ of the Harper map
for $a= 0.3, 0.5$ (SNA) and $a= 0.4, 0.6$ (continuous invariant curves).
Notice that $a= 0.4$ and $a= 0.6$ correspond to
different gaps which are labelled by the indices of the continuous curves 
$8$ and $5$, respectively.   
}
\end{figure}

\begin{figure}
\begin{tabular}{cc}
\hspace{-0.05\linewidth}
\includegraphics[width=0.55\linewidth]{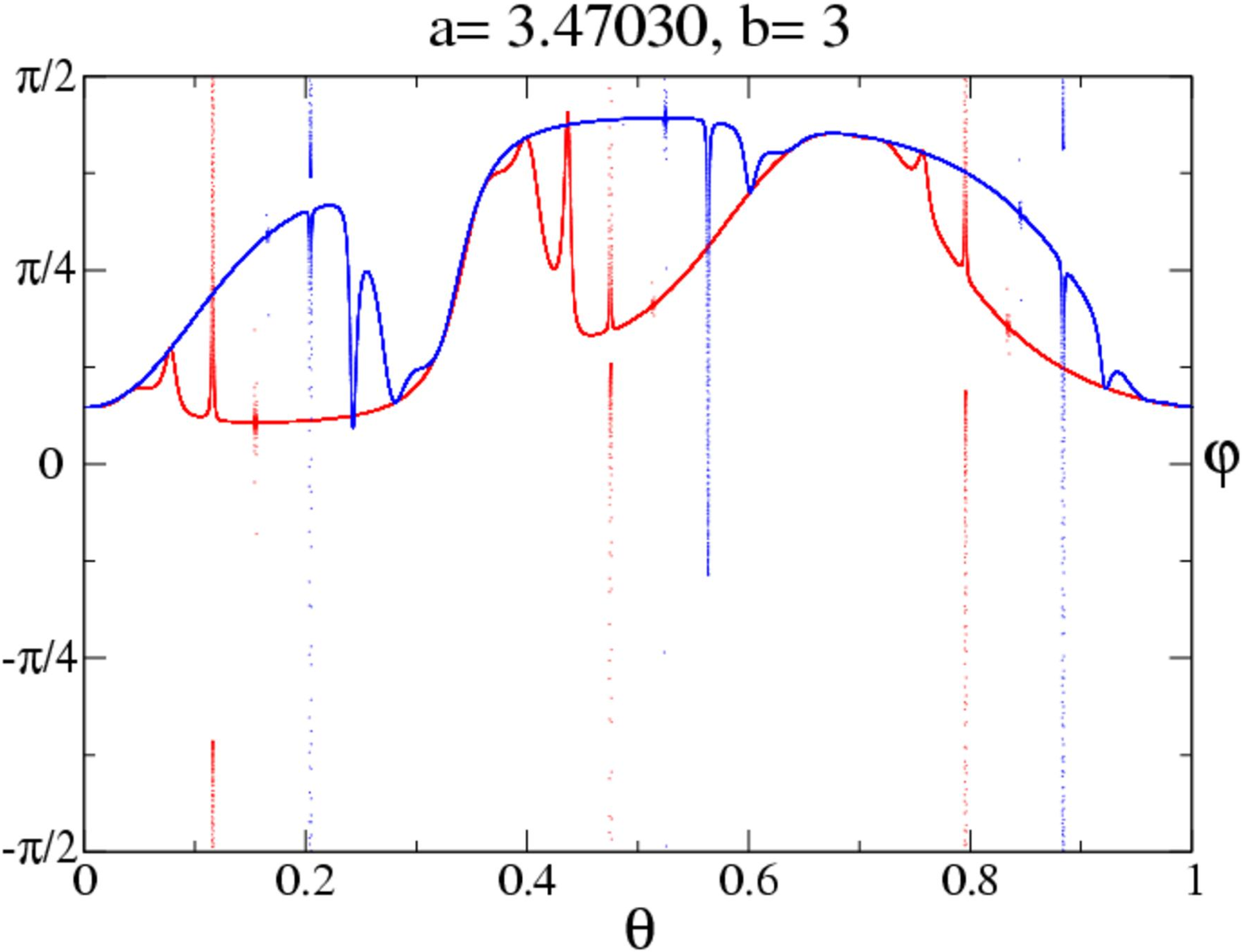} &
\hspace{-0.05\linewidth}
\includegraphics[width=0.55\linewidth]{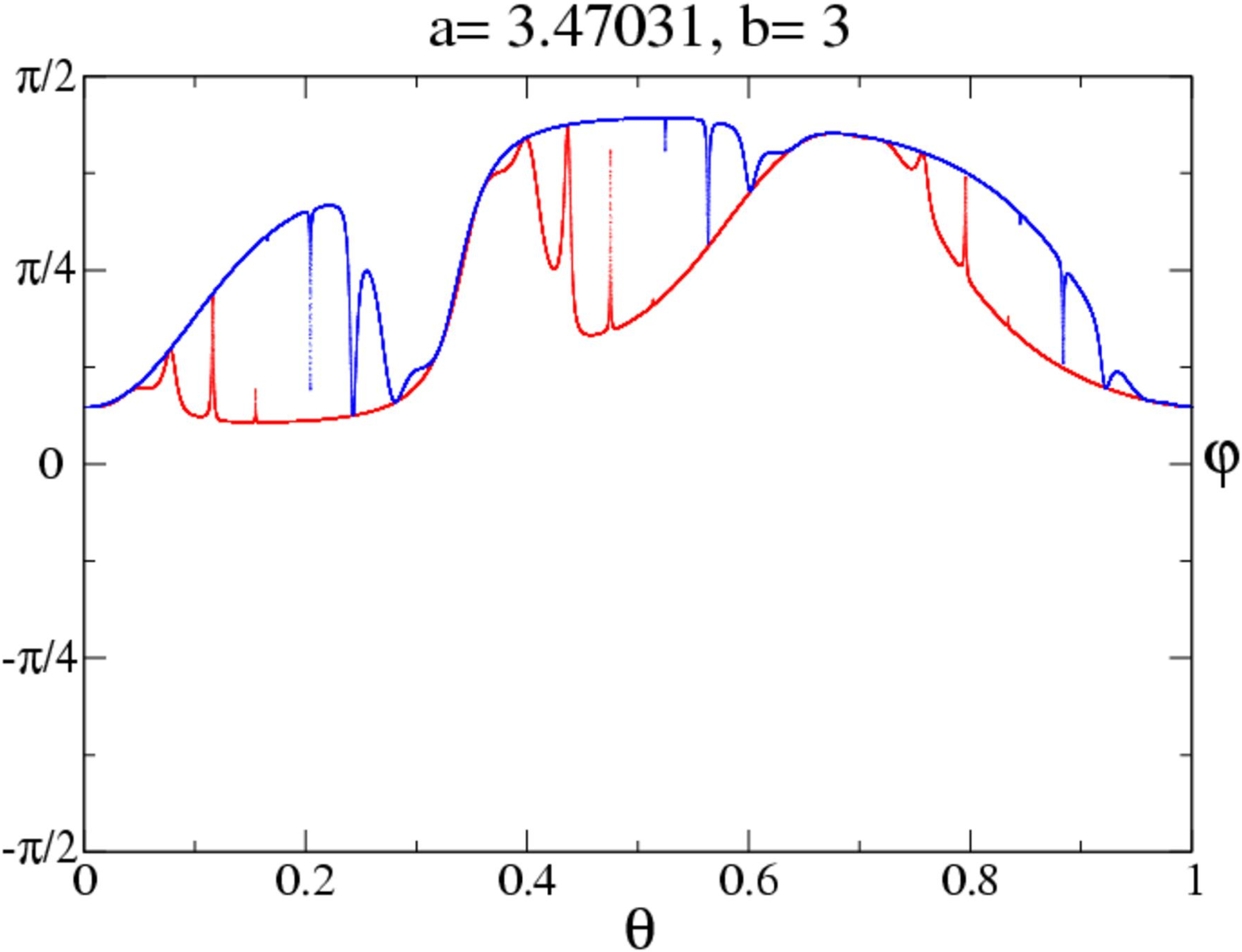} 
\end{tabular}	
\caption{\label{fig:harperb-3b}
The attractor $Y^u$ and the repellor $Y^s$ of the Harper map for 
$a= 3.47030$ (SNA) and $a= 3.47031$ (continuous invariant curves). 
Notice the dramatic 
change in the dynamics with a very small perturbation of $a$. The value 
$a= 3.47031$ belongs to the rightmost gap, whose index is $0$  
(see Figure~\ref{fig:lyapb-3}).
}
\end{figure}

We also selected several values of $a$ and computed 
the attractor $Y^u$ and the repellor $Y^s$ 
of the Harper map for several values of $a.$ The results are displayed in 
Figures~\ref{fig:harperb-3a} and \ref{fig:harperb-3b}.  
Since $y\in{\overline{\mR}}$ represents a slope, 
in the pictures we display 
the angle $\varphi= \tan^{-1} y \in [-\pi/2,\pi/2]$ corresponding 
to such a slope. In this representation we identify 
$\varphi= \pi/2$ with $\varphi= -\pi/2$, 
because a line of slope $+\infty$ is a line with slope $-\infty$. 
The objects shown are graphs of functions 
$y^u(\theta)$, $y^s(\theta)$, respectively, of the slope $y\in{\overline{\mR}}$
with respect to $\theta$.

In this Letter we have shown  the  existence of SNA in Harper maps 
with $\omega$ irrational, $|b|>2$ and $a$ in the spectrum without 
resorting to the possible localization properties of the spectral problem. 
Recall that $a$ is a \emph{point eigenvalue} of a Harper operator 
$H_{b,\omega,\theta_0}$ if the corresponding eigenvalue (Harper) equation
 has a nontrivial \emph{localized} solution 
$\psi=(\psi_n)_{n\in \mZ}$ which is square integrable or even 
\emph{decays exponentially} with $|n|$. 
If the set of localized eigenvectors of a Harper operator forms 
a complete orthogonal basis of $l^2(\mZ)$ then the spectrum is 
\emph{pure-point}.

In previous work on the existence of SNA in Harper maps, localization was 
seen as a justification for the strangeness of SNA, in the regime of 
nonzero Lyapunov exponents 
\rmcite{ketoja-satija,ketoja-satija:self,mestel-osbaldestin-winn,
mestel-osbaldestin:orchid,mestel-osbaldestin:garden}. 
As we have seen, we do not use localization to prove the existence of SNA. 
Besides,  localization may not hold in all the Harper maps 
studied here.
Indeed, an energy $a$ in the spectrum of a Harper operator 
with nonzero Lyapunov exponent 
(for which the Harper map has an SNA) may not be an eigenvalue of the operator.
Indeed, if $\omega$ is not Diophantine, localization may only hold for 
$|b|>2$ large enough (depending on $\omega$) \rmcite{avila-jitomirskaya}.
Even in the Diophantine case,  the spectrum also contains a 
residual set  of energies which are not point eigenvalues  
\rmcite{jitomirskaya-simon,puig:eliasson}. 
Thus, there are SNA  in the family of Harper maps 
without localization for the corresponding 
operator or arithmetic properties on $\omega$.


\def\cprime{$'$}

\end{document}